\documentstyle[prb,aps,preprint]{revtex}

\begin{document}
\def\luni{LuNi$_2$B$_2$C}
\def\yni{YNi$_2$B$_2$C}
\def\<{\langle}
\def\>{\rangle}

\title{Optimal spectrum for the borocarbides \yni\ and \luni}
\author{S. Manalo}
\address{Institut f\"ur Theoretische Physik, J. K. Universit\"at Linz, 
Altenbergerstr. 69, A-4040 Linz, Austria}
\author{E. Schachinger}
\address{Institut f\"ur Theoretische Physik, Technische Universit\"at
Graz, 
Petersgasse 16, A-8010 Graz, Austria}
 
\date{\today}
\maketitle
\begin{abstract}
The concept of an optimal electron-phonon interaction spectral
density as an Einstein spectrum which allows to
describe all physical properties of a superconductor in an
optimal way is developed from Carbotte's original definition of
an optimum spectrum. It is shown, using the borocarbides \yni\ and
\luni\ as examples, that such a concept is meaningful even
for anisotropic systems. An Einstein spectrum is sufficient for
clean-limit systems, a 2$\delta$-peak spectrum is better suited
for anisotropic systems with impurities.
\end{abstract}
\newpage
\section{INTRODUCTION}

Conventional superconductors are well described by Eliashberg
theory\cite{Eli} which treats superconductivity as a
boson-exchange phenomenon. The dominant feature of this theory
is the electron-phonon interaction spectral function
$\alpha^2F(\omega)$ which can be determined from tunneling
experiments\cite{McMill} or theoretically from band structure
calculations. Using such an $\alpha^2F(\omega)$ within Eliashberg
theory allows to reproduce the superconducting properties of a
conventional superconductor within experimental accuracy and
this established the phonons as the exchange boson between the
two charge carriers building the Cooper pair in conventional
superconductors.

Concentrating on isotropic systems Carbotte \cite{Carb} developed
the concept of an optimum spectrum based on earlier work of
Leavens\cite{Leav} and Mitrovi\'c and Carbotte.\cite{Mitr} Such
a spectrum can be developed from a theorem which states that for
a given strength $A = \int_0^\infty d\omega\,\alpha^2F(\omega)$
of the spectral density $\alpha^2F(\omega)$ the best shape that
will maximize the critical temperature $T_c$ is a delta function
spectrum
\begin{equation}
 \alpha^2F(\omega)_{\rm opt} = A\delta[\omega-\omega^\star(\mu^\star)],
\end{equation}
with the delta function placed at the frequency $\omega^\star(\mu^\star)$
at which the functional derivative $\delta T_c/\delta\alpha^2F(\omega)$
displays its maximum for a fixed value of the Coulomb pseudopotential
$\mu^\star$. Carbotte\cite{Carb} extended this concept to encompass
other physical properties such as $2\Delta(0)/k_BT_c$, the zero
temperature gap $\Delta(0)$ to $T_c$ ratio, and a number of others.
This concept establishes that a relation
\begin{equation}
  X = A\,x(\mu^\star)
\end{equation}
always exists, where $X$ stands for $T_c$, $2\Delta(0)/k_BT_c$, etc.
and $x(\mu^\star)$ is a universal number determined from Eliashberg
theory for each property $X$ and which varies only slightly with
$\mu^\star$.

In essence the optimum spectrum gives information about the phonon
frequency important to maximize a certain physical property (such
as $T_c$) of a conventional superconductor. Such a concept is very
appealing and it suggests an expansion to the concept of an {\it optimal}
spectrum which is again a delta peak spectrum with a delta peak of
strength $A$ at some position $\omega^\star(\mu^\star)$ both chosen
to reproduce all known properties of a superconductor optimally.
Such a spectrum will then provide information on the phonon mode
most important for a specific superconductor if an $\alpha^2F(\omega)$
cannot be derived from experiment. It can also help to develop
an $\alpha^2F(\omega)$ in all cases where the phonon density of
states $G(\omega)$ is known.
 
We would like to put this concept to test using the borocarbides
\luni\ and \yni\ for which extensive experimental data exist\cite{Manal}
and for which $G(\omega)$ is known from theoretical
work.\cite{gompf,weber} From experimental data of the upper
critical field $H_{c2}$ which displays a pronounced upward curvature close
to $T_c$ in single crystal\cite{Li} and polycrystalline\cite{Manal}
samples we also assume these systems to be anisotropic.\cite{prosch}
Shulga {\it et al.}\cite{shulga} explained this upward curvature
of $H_{c2}(T)$ close to $T_c$ by considering two bands, one
of which being more deeply involved in the transport properties
of the compound. The authors utilized an $s$-wave
electron-phonon Eliashberg formalism and there is growing
evidence that the order parameter in \yni\ is indeed of
$s$-wave symmetry.\cite{gonn} It is interesting to note
in passing that concept introduced by Prohammer and
Schachinger\cite{prosch} is effectively a two-band model
described by an anisotropic electron-phonon interaction
spectral density.\cite{lang,pitsch}

Sec.~II of this paper reviews the theoretical background, Sec.~III
discusses the results of our analysis, and, finally, in Sec.~IV
our conclusions are drawn.

\section{THEORY}

The theoretical approach towards a theory of anisotropic
polycrystalline superconductors within the framework of
Eliashberg theory is based on the separable model for the
anisotropic electron-phonon interaction introduced by Markovitz and
Kadanoff\cite{marka} which was extended by Daams and Carbotte\cite{daamcar}
to describe an anisotropic electron-phonon interaction spectral function:
\begin{equation}
  \alpha^2F(\omega)_{\bf k, k'} = (1+a_{\bf k})\alpha^2F(\omega)
  (1+a_{\bf k'}), 
\label{spek} 
\end{equation}
where {\bf k} and ${\bf k'}$ are the incoming and outgoing quasi-particle
momentum vectors in the electron-phonon scattering process and $a_{\bf k}$
is an anisotropy function with the important feature
$\langle a_{\bf k}\rangle = 0$, where $\langle\cdots\rangle$ denotes the
Fermi surface average. As anisotropy effects are generally assumed
to be rather small, it is sufficient to keep the mean square anisotropy
$\langle a^2\rangle$ as the important anisotropy parameter. Finally,
$\alpha^2F(\omega)$ is the electron-phonon interaction spectral density
of the equivalent isotropic system.

Thermodynamic properties of a superconductor are calculated from the
free energy difference $\Delta F$ between the normal and superconducting
state:\cite{bard}
\begin{equation}
\Delta F = \pi TN(0)\sum\limits_n^{\omega_c}\left\langle
 \left(\sqrt{\tilde\omega_{\bf k}^2(\omega_n) + 
 \tilde\Delta_{\bf k}^2(\omega_n)} - 
 \left\vert\tilde\omega_{\bf k}(\omega_n)\right\vert\right)
 \left( 1 - {\left\vert\tilde\omega^0_{\bf k}(\omega_n)\right\vert
 \over{\sqrt{\tilde\omega_{\bf k}^2(\omega_n) +
 \tilde\Delta_{\bf k}^2(\omega_n)}}}\right)\right\rangle, 
\label{free-energy}
\end{equation}
with the quasiparticle density of states $N(0)$ at the Fermi level, 
the renormalized quasiparticle frequencies $\tilde\omega_{\bf
k}(\omega_n)$ and the Matsubara gaps $\tilde\Delta_{\bf k}(\omega_n)$
which are the solutions of the nonlinear $s$-wave Eliashberg equations:
\begin{mathletters}
\label{eq:eli}
\begin{eqnarray} 
\tilde\omega_{\bf k}(\omega_n) &=& \omega_n + \pi T \sum\limits_m^{\omega_c}
 \left\langle\left(\lambda_{\bf k,k'}(m-n) +
 \delta_{m,n}{t_{\bf k,k'}^+\over{T}}\right){\tilde\omega_{\bf k'}(\omega_m)
 \over{\sqrt{\tilde\omega_{\bf k'}^2(\omega_m) + \tilde\Delta_{\bf
 k'}^2(\omega_m)}}}\right\rangle' \label{eleqom}\\
 \tilde\Delta_{\bf k}(\omega_n) &=& \pi T \sum\limits_m^{\omega_c}\left\langle
 \left( \lambda_{\bf k,k'}(m-n) - \mu_{\bf k,k'}^\star +
 \delta_{m,n}{t_{\bf k,k'}^+\over{T}}\right){\tilde\Delta_{\bf k'}(\omega_m)
 \over{\sqrt{\tilde\omega_{\bf k'}^2(\omega_m) +
 \tilde\Delta_{\bf k'}^2(\omega_m)}}}\right\rangle'.
\label{eleqdel} 
\end{eqnarray}
\end{mathletters}
The $\omega^0_{\bf k}(\omega_n)$ are the normal state quasiparticle
frequencies determined by
\begin{equation} 
\tilde\omega^0_{\bf k}(\omega_n) = \omega_n + \pi T \sum_m^{\omega_c}
 \left\langle\lambda_{\bf k,k'}(m-n) +
 \delta_{m,n}{t_{\bf k,k'}^+\over{T}}\right\rangle'{\rm sgn}\,\omega_m. 
\end{equation}
In these equations $\omega_c$, the cutoff frequency, is usually an
integer multiple of the Debye frequency of the system,
$\omega_n = \pi T(2n+1), n = 0,\pm1,\pm2,\ldots$,
$t_{\bf k,k'}^+ = 1/(2\pi(\tau_{tr})_{\bf k,k'})$ is the
anisotropic scattering rate due
to inelastic impurity scattering with $(\tau_{tr})_{\bf k,k'}$ as
the anisotropic transport relaxation time, $\mu^\star_{\bf k,k'}$
is the anisotropic Coulomb pseudopotential, and
\begin{equation}
\lambda_{\bf k,k'}(m-n) = 2\int\limits_0^\infty\! d\Omega\, 
{\Omega\,\alpha^2F(\Omega)_{\bf k,k'}\over{\Omega^2 +
(\omega_m-\omega_n)^2}}.
\label{lambdak}
\end{equation}

In case of weak anisotropy effects the ${\bf k,k'}$ dependence of
the Coulomb pseudopotential and of the impurity scattering 
is neglected and the anisotropy of the Matsubara
gaps is described by the ansatz
\begin{equation}
\tilde\Delta_{\bf k}(\omega_n) = \tilde\Delta_0(\omega_n) + a_{\bf
k}\tilde\Delta_1(\omega_n), 
\label{delsep}
\end{equation}
with $\tilde\Delta_{0,1}(\omega_n)$ being isotropic functions.
In applying equation (\ref{delsep}) 
to Eqs.~(\ref{eq:eli}) only terms of the order of $\< a^2\>$
are kept.

The upper critical field $H_{c2}(T)$ of an anisotropic polycrystalline
superconductor employs, in addition, a separable ansatz to describe
the anisotropy of the Fermi velocity $v_{F,{\bf k}}$\cite{prosch}
\begin{equation}
v_{F,{\bf k}} = (1 + b_{\bf k})\langle v_F\rangle,
\label{vf}
\end{equation}
with $\langle v_F\rangle$ the isotropic Fermi velocity. $b_{\bf k}$ is an
anisotropy function defined in the same way as $a_{\bf k}$.
Again, only terms of the order $\langle b^2\rangle$ are kept
in case of small anisotropy effects.
The upper critical field in its temperature dependence is then
described by the following set of equations:\cite{prosch}
\begin{mathletters}
\label{hc2}
\begin{eqnarray}
\tilde\Delta_{\bf k}(\omega_n) &=& 
\pi T\sum_m(1+a_{\bf k})\lambda(m-n)\left\langle (1+a_{\bf k'})
\tilde\Delta_{\bf k'}(\omega_m)\chi_{\bf k'}(m)\right\rangle'\nonumber\\
&& -\pi T\sum\limits_m\left(\mu^\star -\delta_{n,m} {t^+\over{T}}\right)
\left\langle \tilde\Delta_{\bf k'}(\omega_m)\chi_{\bf k'}(m)\right\rangle',\\
 \label{gapans2}
\chi_{\bf k}(n) &=& {2\over{\sqrt{\alpha_{\bf k}}}}\int\limits_0^\infty\!dx\, 
{\rm e}^{-x^2} \tan^{-1}\left({\sqrt{\alpha_{\bf k}}x
\over{\left\vert\tilde\omega_{\bf k}(\omega_n)\right\vert}}
 \right), \label{chians2} 
\end{eqnarray}
\end{mathletters}
 and
\begin{equation}
\alpha_{\bf k} = {e\over{2}}H_{c2}(T)(1+b_{\bf k})^2\langle v_F\rangle^2.
\label{alphak}
\end{equation}

\section{DATA ANALYSIS}
\subsection{The $\alpha^2F(\omega)$ spectra}

As there are no data available which would allow to determine the
electron-phonon interaction spectral density $\alpha^2F(\omega)$
directly by inversion, we have to start our analysis using the
phonon density of states $G(\omega)$ which is known from theoretical
work.\cite{gompf,weber} We construct the electron-phonon
interaction spectral density using the ansatz
\begin{equation}
\alpha^2F(\omega) = c\omega^sG(\omega),
\end{equation}
with constants $c$ and $s$. The coupling strength between
electrons and phonons is generally assumed to become weaker with
increasing energy $\omega$ which leaves $s$ to be negative and
it is set to $-1/2$ for simplicity, a value which was also
proposed by Junod {\it et al.}\cite{junod} in their analysis of
A15-compounds. (Gonnelli {\it et al.}\cite{gonn} used instead a
two step weighting function which was determined by a fit to the
high temperature part of the resistivity of a very pure single
crystal \yni\ sample.) The constant $c$ is then used to rescale the
spectrum to obtain the experimentally observed $T_c$
($15.45\,$K in the case of \yni\ and $16\,$K for
\luni) using the linearized versions of Eqs.~(\ref{eq:eli})
applied to an isotropic system. Fig.~\ref{b_spek} presents
the low energy part of the resulting $\alpha^2F(\omega)$
obtained for a Coulomb pseudopotential $\mu^\star = 0.13$.
For \yni\ $\lambda = \lambda(0) = 1.071$ (solid
squares) and for \luni\ $\lambda = 1.267$ (solid
triangles) which identifies both materials as medium coupling
strength superconductors.

In the analysis presented here, the actual energy dependence of
the $\alpha^2F(\omega)$ spectrum is of no importance as we are
going to replace the $\alpha^2F(\omega)$ by an Einstein spectrum
with its $\delta$-peak of strength $A$ at some fixed frequency
$\omega^\star(\mu^\star)$ with $A$ and $\omega^\star(\mu^\star)$
chosen to give the measured $T_c$, the appropriate value for
$\lambda$, and the best possible fit to experiment.
Nevertheless, these $\alpha^2F(\omega)$ spectra already give 
a pretty good idea where to place the $\delta$-peak because
a quite natural choice is to place the $\delta$-peak at the
center of the area under the respective $\alpha^2F(\omega)$
spectrum.

\subsection{\yni, clean-limit case}

We start the procedure using data for the system \yni\ and begin with
the temperature dependence of the thermodynamic critical field
\begin{equation}
  \mu_0 H_c(T) = \sqrt{2\mu_0 \Delta F(T)},
\end{equation}
with $\mu_0$ the permeability constant of vacuum, and with the
deviation function
\begin{equation}
  D(t) = {H_c(T)\over H_c(0)}- (1-t^2),
\end{equation}
where $t = T/T_c$. The free energy difference $\Delta F(T)$ is
determined experimentally by a twofold integration of the specific heat
\begin{equation}
\Delta F(T) = -\intop_0^T\intop_0^{T'}{\rm d}T'{\rm d}T''{\Delta
C(T'')\over{T''}}
\end{equation} 
measured in magnetic fields between $0\le \mu_0 H\le
9\,$T.\cite{Manal,michor}

The best agreement is found for $\omega^\star(\mu^\star) = 17\,$meV
which is close to the center of the area under the $\alpha^2F(\omega)$
spectrum (solid squares, Fig.~\ref{b_spek}) and a $\lambda = 1.054$.
The numerical
results obtained from solving Eqs.~(\ref{eq:eli}) together with
(\ref{free-energy}) are presented in Figs.~\ref{y_delta_hc} and
\ref{y_delta_dev} for $H_c(T)$ and $D(t)$ respectively. (We notice
the large error bar on the $D(t)$ data points which results from
the method used to determine $\Delta F(T)$ from experiment and
from the necessary extrapolation of the data to $T\to 0$ to
find $H_c(0)$.) Experimental data are shown as open squares,
theoretical results are presented for the isotropic case
($\langle a^2\rangle = 0$, solid line), $\langle a^2\rangle = 0.01$
(dashed line), $\langle a^2\rangle = 0.02$ (dotted line), and
$\langle a^2\rangle = 0.03$ (dash-dotted line). It is obvious
that results for $0.02\le\langle a^2\rangle\le 0.03$ agree
best with experiment.

In the next step data for the upper critical field $H_{c2}(T)$
is used to determine the anisotropy parameter $\langle b^2\rangle$
of the Fermi velocity $\langle v_F\rangle$ within the above range of
$\langle a^2\rangle$ values. Here, the upward curvature of $H_{c2}(T)$
at $T_c$ is used to fit $\langle b^2\rangle$ while
$\langle v_F\rangle$ sets the scale according to Eq.~(\ref{alphak}).
In order to remove ambiguities, an experimental value of the average
Fermi velocity $\langle v_F\rangle$ can be derived from the
plasma frequency $\Omega_p$ by using
\begin{equation}
\hbar\Omega_p = \sqrt{4\pi e^2\langle v_F\rangle N(0)/3}.
\end{equation}
For \luni\ $\Omega_p = 4\,$meV\cite{bommel} which results in
$\langle v_F\rangle \simeq 0.28\times 10^6\,$m/s. Application
of Eqs.~(\ref{hc2}) with $\langle b^2\rangle = 0.315$ and
$\langle v_F\rangle = 0.275\times 10^6\,$m/s results in the
dashed curve in Fig.~\ref{y_2delta_hc2} for $\langle a^2\rangle = 0.02$
while $\langle b^2\rangle = 0.305$ and
$\langle v_F\rangle = 0.285\times 10^6\,$m/s are found for
$\langle a^2\rangle = 0.03$ (solid line, Fig.~\ref{y_delta_hc2})
to give an optimal fit to experiment (open squares).

The same procedure can be applied to analyze the experimental data
of \luni\ in a clean limit approach using a single peak Einstein
spectrum. The peak is set at the energy $\omega^\star(\mu^\star) =
15\,$meV, again close to the center of area under the $\alpha^2F(\omega)$
spectrum (solid triangles, Fig.~\ref{b_spek}). The results of such a
calculation are presented in Table~\ref{t1} and compared to the
\yni\ anisotropy parameters.

This proves that it is indeed possible to describe the features of
\yni\ and \luni\ rather well using Eliashberg theory for anisotropic
$s$-wave superconductors and a simple Einstein spectrum to describe the
energy dependence of the electron-phonon interaction spectral
density. Nevertheless, the agreement at low temperatures is
still not perfect (see insets of Figs.~\ref{y_delta_hc} and \ref{y_delta_hc2})
which demonstrate that the low temperature data are somewhat overestimated by
our analysis.

\subsection{Impurity scattering}

In reality, the sample develops some residual resistivity at low
temperatures which is an indication of some impurity content.
Thus, a clean limit analysis of experimental data as it was
presented in the previous subsection can only be a first step
which allows to put some margins on the various anisotropy
parameters. It would then be standard procedure\cite{webers} to
load the sample under investigation in a controlled way
with some impurities and to measure the change in $T_c$ and in the
residual resistivity $\rho_n$ as a function of impurity
concentration. This gives another, rather reliable estimate
for the anisotropy parameter $\langle a^2\rangle$\cite{marka} and allows
to calculate the impurity parameter $t^+$ which enters
Eqs.~(\ref{eq:eli}) and (\ref{hc2}) from the Drude relation
\begin{equation}
   t^+ = {\rho_n\hbar\Omega_p\over 8\pi^2}.
\end{equation}

Such data is not available for \yni\ and \luni\ and we have to develop
a different strategy to achieve a more realistic simulation. We make
use of the fact that the critical temperature of an anisotropic
clean limit system $(t^+ = 0)$, $T_{c0}$, is always greater than
the critical temperature of a realistic system with $t^+ > 0$.
One can therefore choose a hypothetical value for $T_{c0}$ and
calculate how $T_c$ decreases with increasing values of $t^+$ for
a fixed value of $\langle a^2\rangle$. Results of such a
calculation are shown in Fig.~\ref{ytctplus}
(solid line for $T_{c0} = 15.5\,$K and $\langle a^2\rangle = 0.03$,
dashed line for $T_{c0} = 15.5\,$K and $\langle a^2\rangle = 0.035$,
dotted line for $T_{c0} = 15.55\,$K and $\langle a^2\rangle = 0.035$, and
dash-dotted line for $T_{c0} = 15.6\,$K and $\langle a^2\rangle = 0.04$).
This defines
the value $t^+$ necessary in the simulation for the realistic
system to obtain the experimental value of the critical temperature
(labeled $T_c$, \yni\ in Fig.~\ref{ytctplus}).

We also have to keep in mind that adding impurities increases
$H_{c2}(T)$ even in isotropic systems.\cite{prosch,werth}
Moreover, adding impurities `smears out'
the anisotropy\cite{marka} which results in increasing
values of $H_c(T)$ and an additional increase of $H_{c2}(T)$,
and, furthermore, in a
less pronounced upward curvature of $H_{c2}(T)$ close to $T_c$%
\cite{webers} if $\langle a^2\rangle$ and $\langle b^2\rangle$
are kept constant in the calculations. Thus, one will have to
compensate for adding impurities by an increase of the anisotropy
parameters. On the other hand, the anisotropy parameters found
for the clean limit system already seem to be rather realistic
(they are close to the values given by Manalo {\it et al.}\cite{Manal}
found by a more elaborate analysis) and we will therefore relay
to other means to reestablish agreement with experiment. 

Adding a second $\delta$-peak to the
$\alpha^2F(\omega)$ spectrum with its position and strength
chosen that, again, the best possible agreement with experiment
can be established by changing the anisotropy parameters only
minimally is suggested by functional derivatives
$\delta H_{c2}(T)/\delta\alpha^2F(\omega)$.\cite{mars} In
particular, functional derivatives reveal that adding
spectral weight at high energies to the $\alpha^2F(\omega)$
spectrum makes it less effective for $H_{c2}(T)$ at low
temperatures while adding spectral weight at low energies
has just the opposite effect. At high temperatures $H_{c2}(T)$
is far less sensitive to changes in the spectral weight. This
is of importance as our calculations overestimate $H_{c2}(T)$
at low temperatures already for the clean limit system in the
case of \yni\ (Fig.~\ref{y_delta_hc2}). We therefore add some
spectral weight at higher energies and
a natural choice for the energy $\omega_2$ at which
this second $\delta$-peak is to be placed is the energy of the maximum
in the $\alpha^2F(\omega)$ spectrum, i.e. $\omega_2 = 21\,$meV
for \yni\ (Fig.~\ref{b_spek}). The strength $A_2$ of this
second peak is chosen to be a tenth of the strength $A$ of the
primary $\delta$-peak. This new 2$\delta$-peak spectrum has
then to be rescaled to reproduce the experimental value of
$T_c$ for the fixed value $\mu^\star = 0.13$. This procedure
results in a $\lambda = 1.04$ just marginally smaller than
the $\lambda$ of the Einstein spectrum.

Using this new $2\delta$-spectrum $H_c(T)$ and $H_{c2}(T)$ are
recalculated
with the parameters shown in Table~\ref{t2}. The results
for $H_c(T)$ are presented in Fig.~\ref{y_thermo}. We see
that the low temperature values of $H_c(T)$ are still slightly
overestimated by our model calculations, and the results for
$H_{c2}(T)$ (Fig.~\ref{y_2delta_hc2}) reveal that $H_{c2}(T)$ is
now a bit underestimated at very low temperatures but
otherwise the agreement is almost perfect for all sets of
impurity and anisotropy parameters.

Again, the same procedure can be applied to the system \luni.
We use a 2$\delta$-spectrum with the second peak placed at the
maximum in the $\alpha^2F(\omega)$ spectrum (solid triangles,
Fig.~\ref{b_spek}) $\omega_2 = 10\,$meV and with its strength
given by $A/A_2 \simeq 7.14$ because $H_{c2}(T)$ was originally
slightly underestimated. The resulting $\lambda = 1.237$ and
the anisotropy parameters for best agreement with experimental
data found for this $2\delta$-spectrum are quoted in Table~\ref{t2}.

It is important to emphasize at this point that the agreement
between theoretical predictions and experiment is equally
good for all sets of parameters quoted in Table~\ref{t2}. Thus,
only an experiment in which the residual resistivity of the
sample is measured can finally help in pinning down the `real'
anisotropy parameters. Our analysis only helped to substantially
narrow the range of realistic values of the anisotropy
parameters.

\section{SUMMARY AND CONCLUSION}

We investigated two borocarbide systems, namely \yni\ and \luni,
to prove whether the concept of an optimal electron-phonon
interaction spectrum is applicable to anisotropic superconductors
in general and to the borocarbides in particular. Such a concept
seems to be very helpful if only little is known about the
electron-phonon interaction spectral function $\alpha^2F(\omega)$.

This concept is indeed
applicable to anisotropic clean limit systems where an Einstein spectrum
with its peak placed near the center of the area under a model-%
$\alpha^2F(\omega)$ spectrum proved sufficient to obtain an
excellent agreement between theoretical predictions and experiment
over the whole temperature range for thermodynamics and
$H_{c2}(T)$. The latter is particularly sensitive to anisotropy
effects and details in $\alpha^2F(\omega)$. Nevertheless,
even for this property excellent agreement could be achieved.
This proves that the model of an anisotropic electron-phonon
interaction spectral function $\alpha^2F(\omega)_{{\bf k},{\bf k}'}$
can be used to explain the upward curvature of $H_{c2}(T)$ close
to $T_c$. 

Systems with impurities require compensation of the smearing out
of anisotropy by increasing the values of the anisotropy
parameters $\langle a^2\rangle$ and $\langle b^2\rangle$ if the
Einstein spectrum concept is to be extended even to this case.
This could result in rather big values for these parameters and
in unrealistic values of $\langle v_F\rangle$ necessary to
reproduce $H_{c2}(T)$ on an absolute scale. As a way out of
this problem we offer to allow the optimal spectrum
to be a 2$\delta$-spectrum with the position of the main peak
determined from clean limit calculations. Our analysis
suggests various critical temperatures $T_{c0}$ for the clean-limit
system which is the `origin' of the realistic sample. Measuring
the residual resistivity of the sample under investigation
will then determine $t^+$ which in turn gives the
appropriate anisotropy parameters (see Table~\ref{t2}).

Finally, the applicability of the concept of an optimal
$\alpha^2F(\omega)$ spectrum to the systems \yni\ and \luni\
proved that both are classical $s$-wave electron-phonon
superconductors adequately described by an Eliashberg theory
of anisotropic superconductors.

\section*{Acknowledgments}

The authors want to thank Dr.~H.~Michor and Prof.~G.~Hilscher 
for many fruitful discussions regarding the borocarbides. One of us 
is also very grateful for their support in obtaining the experimental 
data used in this work as a part of her Diploma thesis.\cite{Manal}

\begin{table}
\caption{Anisotropy parameters for \yni\ and \luni\ from the clean limit
calculations. $\omega^\star(\mu^\star)$ is in meV, $T_c$ in K, and
$\langle v_F\rangle$ in $10^6\,$m/s.
}
\begin{tabular}{ccccccc}
Material & $\omega^\star(\mu^\star)$ &  $\lambda$ & $T_c$ & $\langle a^2\rangle$ &
 $\langle b^2 \rangle$ & $\langle v_F\rangle$\\
\hline
 \yni & 17.0 & 1.054 & 15.45 & 0.02 & 0.315 & 0.275\\
      & 17.0 & 1.054 & 15.45 & 0.03 & 0.305 & 0.285\\
\hline
 \luni & 15.0 & 1.174 & 16.0 & 0.02 & 0.255 & 0.288\\
       & 15.0 & 1.174 & 16.0 & 0.03 & 0.25 & 0.298
\end{tabular}
\label{t1}
\end{table}
\begin{table}
\caption{Anisotropy parameters for an anisotropic system with
impurities used to simulate the experimental data found for
\yni. $t^+$ is given in meV and $\langle v_F\rangle$ in
$10^6\,$m/s.}
\label{t2}
\begin{tabular}{cccccc}
  Material & $T_{c0}$ & $\langle a^2\rangle$ & $t^+$ & $\langle b^2\rangle$ &
  $\langle v_F\rangle$\\
\hline
  \yni & 15.5 & 0.03 & 0.366 & 0.330 & 0.288\\
  & 15.5 & 0.035 & 0.306 & 0.325 & 0.293\\
  & 15.55 & 0.035 & 0.684 & 0.345 & 0.295\\
  & 15.6 & 0.04 & 0.961 & 0.355 & 0.301\\
\hline
  \luni & 16.05 & 0.03 & 0.533 & 0.275 & 0.305\\
   & 16.05 & 0.035 & 0.455 & 0.275 & 0.31\\
   & 16.1 & 0.035 & 1.025 & 0.375 & 0.318\\
   & 16.1 & 0.04 & 0.859 & 0.295 & 0.32
\end{tabular}
\end{table}
\begin{figure}
\caption{Spectral densities $\alpha^2F(\omega)$ of \luni\ (solid
triangles) and \yni\ (solid squares) in
the low-energy range, rescaled to obtain the measured critical
temperature $T_c$ for a fixed value of the Coulomb pseudopotential
$\mu^*=0.13$ from the solutions of linearized
Eliashberg equations fro isotropic systems.
}
\label{b_spek}
\end{figure}

\begin{figure}
\caption{The temperature dependence of the thermodynamic critical
field $H_c(T)$ in \yni, obtained from solutions of the Eliashberg
equations
(\ref{eq:eli}) for an Einstein spectrum with its $\delta$-peak
at $\omega^\star(\mu^\star) = 17$\,meV and for various
anisotropy parameters $\langle a^2\rangle$, namely $\langle a^2\rangle = 0$
(isotropic case, solid line), $\langle a^2\rangle = 0.01$ (dashed line),
$\langle a^2\rangle = 0.02$ (dotted line), and $\langle a^2\rangle = 0.03$
(dash-dotted line). The open squares represent experimental data.
}         
\label{y_delta_hc}
\end{figure}

\begin{figure}
\caption{The thermodynamic critical field deviation function $D(t)$
as a function of the reduced temperature $t = T/T_c$
in \yni, obtained from solutions of the Eliashberg equations
(\ref{eq:eli}) for an Einstein spectrum with its $\delta$-peak
at $\omega^\star(\mu^\star) = 17$\,meV and for various
anisotropy parameters $\langle a^2\rangle$, namely $\langle a^2\rangle = 0$
(isotropic case, solid line), $\langle a^2\rangle = 0.01$ (dashed line),
$\langle a^2\rangle = 0.02$ (dotted line), and $\langle a^2\rangle = 0.03$
(dash-dotted line). The open squares represent experimental data.
}         
\label{y_delta_dev}
\end{figure}

\begin{figure}
\caption{The temperature dependence of the upper critical field
$H_{c2}(T)$ in \yni\ obtained from solutions of Eqs.~\ref{hc2}
for an Einstein spectrum with its $\delta$-peak at
$\omega^\star(\mu^\star) = 17$\,meV and for various
anisotropy parameters, namely $\langle a^2\rangle = 0.02$,
$\langle b^2\rangle = 0.315$, and $\langle v_F\rangle =
0.275\times10^6\,$%
m/s (dashed line), and $\langle a^2\rangle = 0.03$,
$\langle b^2\rangle = 0.305$, and $\langle v_F\rangle =
0.285\times10^6\,$%
m/s (solid line). The open squares without visible error bars represent
experimental data obtained from specific heat measurements by entropy
conservation. The two low temperature data points were determined from
resistivity measurements, with error bars indicating the width of the
transition.\cite{Manal}}
\label{y_delta_hc2}
\end{figure}
\begin{figure}
\caption{The critical temperature $T_c$ of an anisotropic superconductor
as a function of $t^+$ which is proportional to the impurity concentration
for fixed values of the anisotropy parameter $\langle a^2\rangle$.
The solid line is for $T_{c0} = 15.5\,$K and $\langle a^2\rangle = 0.03$,
the dashed one for $T_{c0} = 15.5\,$K and $\langle a^2\rangle = 0.035$,
the dotted one for $T_{c0} = 15.55\,$K and $\langle a^2\rangle = 0.035$, and
the dash-dotted one for $T_{c0} = 15.6\,$K and $\langle a^2\rangle = 0.04$.
The thin straight line labeles $T_c$ and indicates the experimental
value of the critical temperature of our \yni\ sample.}
\label{ytctplus}
\end{figure}
\begin{figure}
\caption{The temperature dependence of the thermodynamic critical
field $H_c(T)$ in \yni, obtained from solutions of the Eliashberg
equations
(\ref{eq:eli}) for a 2$\delta$-spectrum for the various
anisotropy parameters of Table~\ref{t2}. The solid line is for
$T_{c0} = 15.5\,$K and $\langle a^2\rangle = 0.03$, dashed is for
$T_{c0} = 15.5\,$K and $\langle a^2\rangle = 0.035$, dotted is for
$T_{c0} = 15.55\,$K and $\langle a^2\rangle = 0.035$, and dash-dotted is
for $T_{c0} = 15.6\,$K and $\langle a^2\rangle = 0.04$.
The open squares represent experimental data.
}         
\label{y_thermo}
\end{figure}
\begin{figure}
\caption{The temperature dependence of the upper critical
field $H_{c2}(T)$ in \yni, obtained from solutions of
equations
(\ref{hc2}) for a 2$\delta$-spectrum for the various
anisotropy parameters of Table~\ref{t2}. The solid line is for
$T_{c0} = 15.5\,$K and $\langle a^2\rangle = 0.03$, dashed is for
$T_{c0} = 15.5\,$K and $\langle a^2\rangle = 0.035$, dotted is for
$T_{c0} = 15.55\,$K and $\langle a^2\rangle = 0.035$, and dash-dotted is for 
$T_{c0} = 15.6\,$K and $\langle a^2\rangle = 0.04$.
The open squares without visible error bars represent
experimental data obtained from specific heat measurements by entropy
conservation. The two low temperature data points were determined from
resistivity measurements, with error bars indicating the width of the
transition.\cite{Manal}
}         
\label{y_2delta_hc2}
\end{figure}


\begin{references}
\bibitem{Eli}G.M.~Eliashberg, Sov. Phys. JETP {\bf 11}, 696 (1960).
\bibitem{McMill}W.L.~McMillan and J.M.~Rowell, \prl {\bf 14}, 108
(1965).
\bibitem{Carb}J.P.~Carbotte, Rev.\ Mod.\ Phys.\ {\bf 62}, 1027 (1990).
\bibitem{Leav}C.R.~Leavens, Solid State Commun. {\bf 17}, 1499 (1975).
\bibitem{Mitr}B.~Mitrovi\'c and J.P.~Carbotte, Solid State Commmun.
{\bf 40}, 249 (1981).
\bibitem{Manal}S.~Manalo, Diploma thesis, Technische
Universit\"at Wien, Karlsplatz 13, 1040 Wien, Austria (1999) unpulished;
S.~Manalo, H.~Michor, M.~El-Hagary, G.~Hilscher, and E.~Schachinger,
\prb (in print) and cond-mat/9911305.
\bibitem{gompf}F. Gompf, W. Reichardt, H. Schober, B. Renker, and 
M. Buchgeister, Phys. Rev. B {\bf 55}, 9058 (1997) 
\bibitem{weber}W. Weber, Universit\"at Dortmund, Institut f\"ur
Theoretische
Physik II, Otto-Hahn-Str. 4, D-44221 Dortmund (unpublished).
\bibitem{Li}Shi Li, {\it et al.}, Int.\ J.\ Mod.\ Phys. {\bf 13},
3725 (1999).
\bibitem{prosch}M. Prohammer and E. Schachinger, {Phys. Rev. B {\bf 36},
8353} (1987).
\bibitem{shulga}V.~Shulga, S.-L.~Drechsler, G.~Fuchs, k.-H.~M\"uller,
K.~Winzer, M.~Heinecke, and K.~Krug, \prl {\bf 80}, 1730 (1998).
\bibitem{gonn}R.S.~Gonnelli, A.~Morello,G.A.~Ummarino, V.A.~Stepanov,
G.~Behr, G.~Graw, V.~Shulga, and S.-L.~Drechsler,
cond-mat/0007033 (unbublished).
\bibitem{lang}E.~Langmann, \prb {\bf 46}, 9104 (1992).
\bibitem{pitsch}W.~Pitscheneder and E.~Schachinger, \prb {\bf 47}, 3300
(1993).
\bibitem{marka}D. Markovitz, L.P. Kadanoff, {Phys. Rev. {\bf 131}, 563}
(1963).
\bibitem{daamcar}J.M. Daams, J.P. Carbotte, J. Low Temp. Phys. {\bf 
43}, 263 (1981).
\bibitem{bard}J. Bardeen and M. Stephen, {Phys. Rev. {\bf 136},
1485} (1964).
\bibitem{junod}A. Junod, T. Jarlborg, and J. Muller, Phys. Rev. B {\bf 
27}, 1568 (1983)
\bibitem{michor}H.~Michor, T.~Holubar, C.~Dusek, and G.~Hilscher, \prb
{\bf 52}, 16165 (1995)
\bibitem{bommel}F. Bommeli, L. Degiorgi, P. Wachter, B.K. Cho,
P.C. Canfield, R. Chau and M.B. Maple, {Phys. Rev. Lett. {\bf 78},
547} (1997)
\bibitem{webers}H.W. Weber, E. Seidl, C. Laa, E. Schachinger,
M. Prohammer, A. Junod, D. Eckert, Phys. Rev. B {\bf 44}, 7585 (1991)
\bibitem{werth}N.R. Werthammer, E. Helfand, and P.C. Hohenberg,
Phys. Rev. {\bf 147}, 295 (1966).
\bibitem{mars}F. Marsiglio, M. Schossmann, E. Schachinger, and
J.P. Carbotte, \prb {\bf 35}, 3226 (1987).

\end{references}
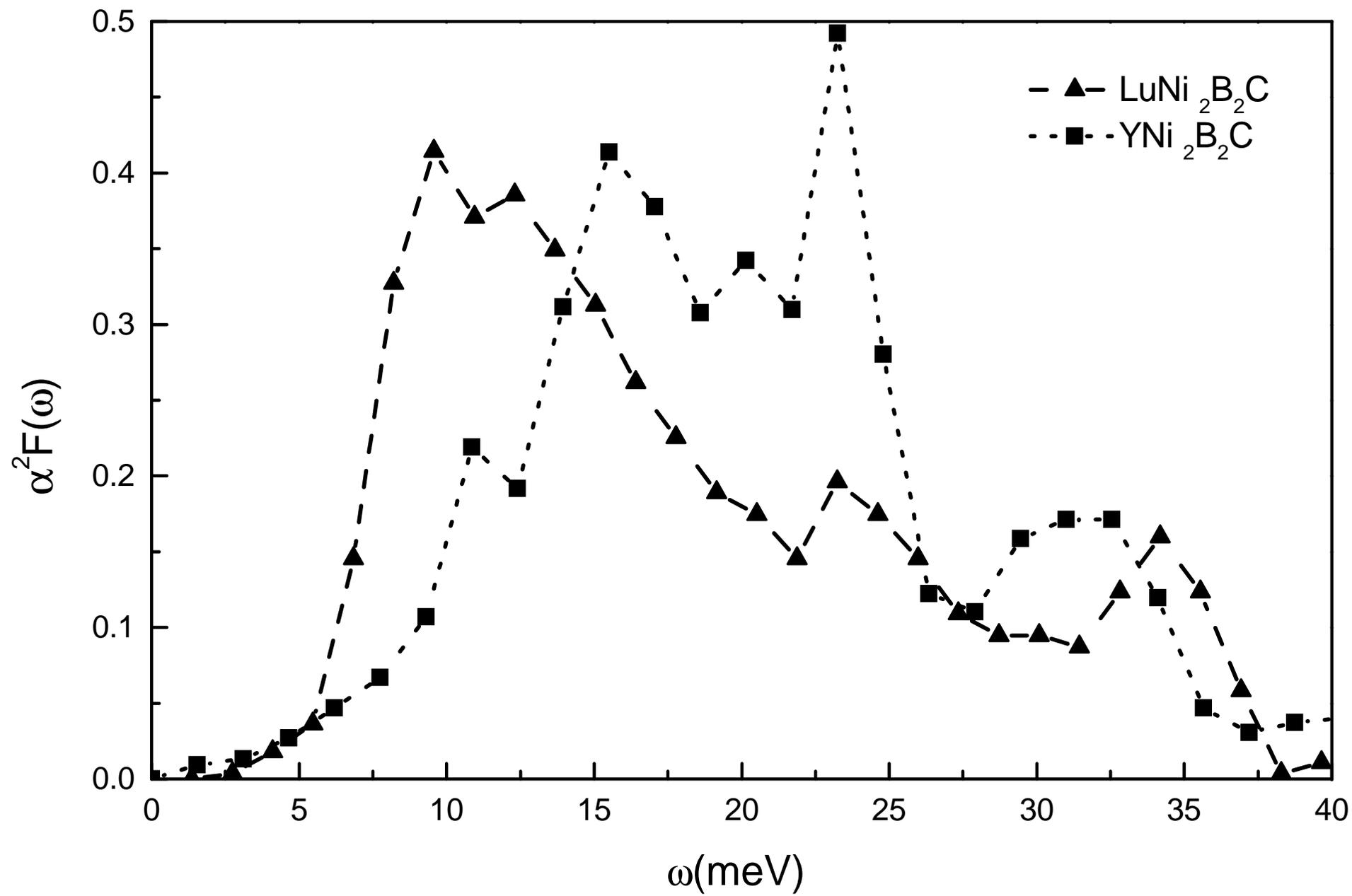
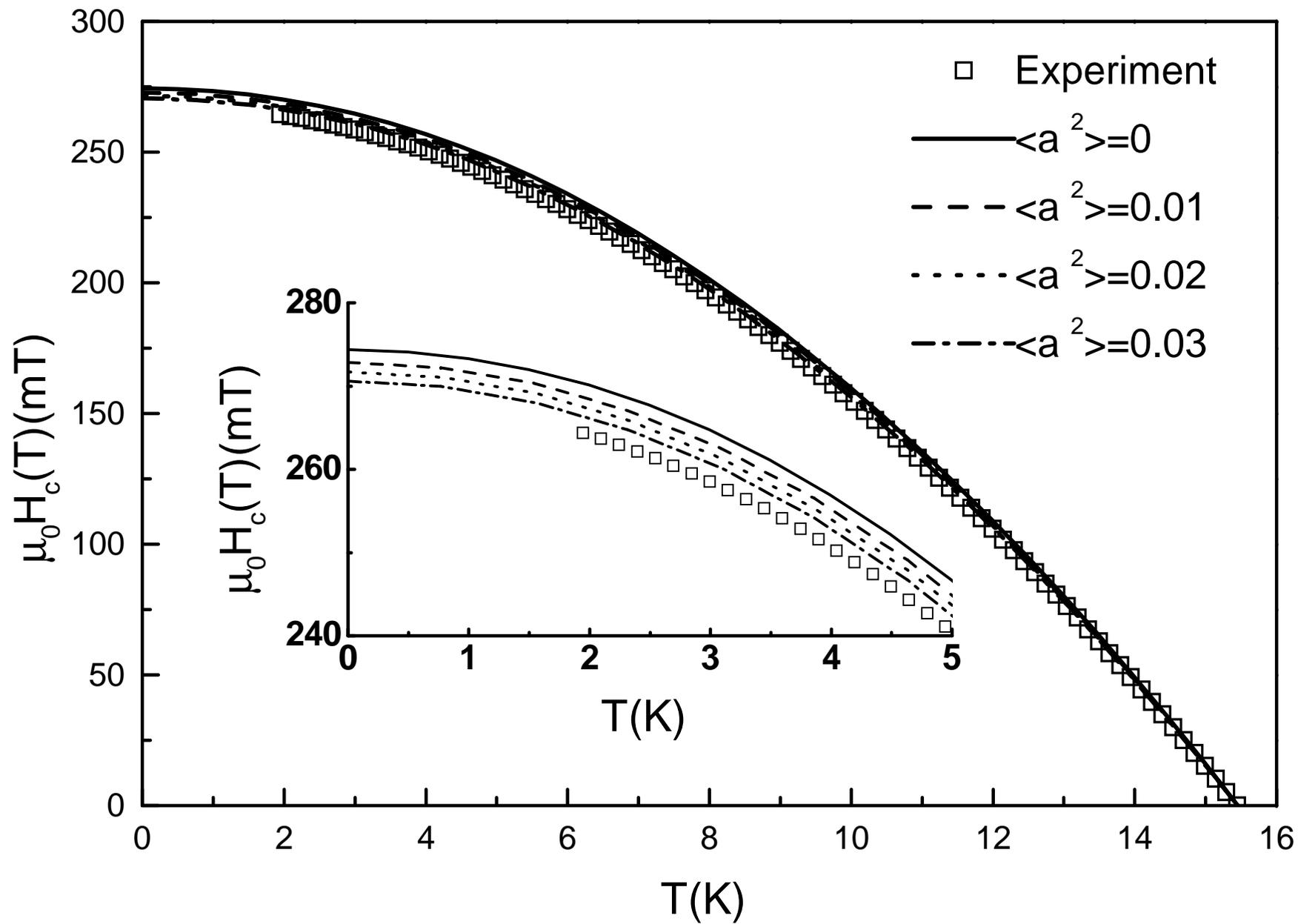
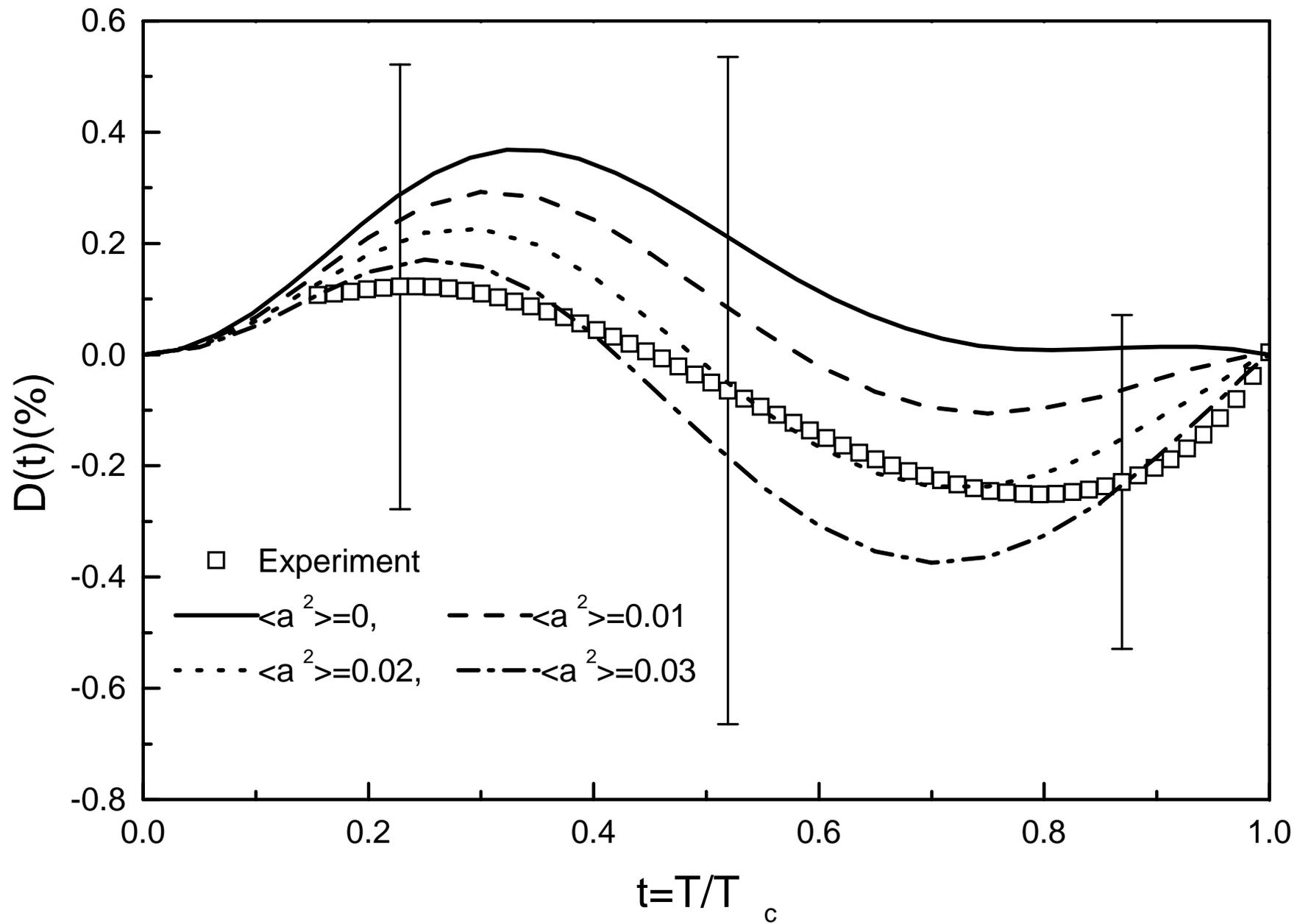
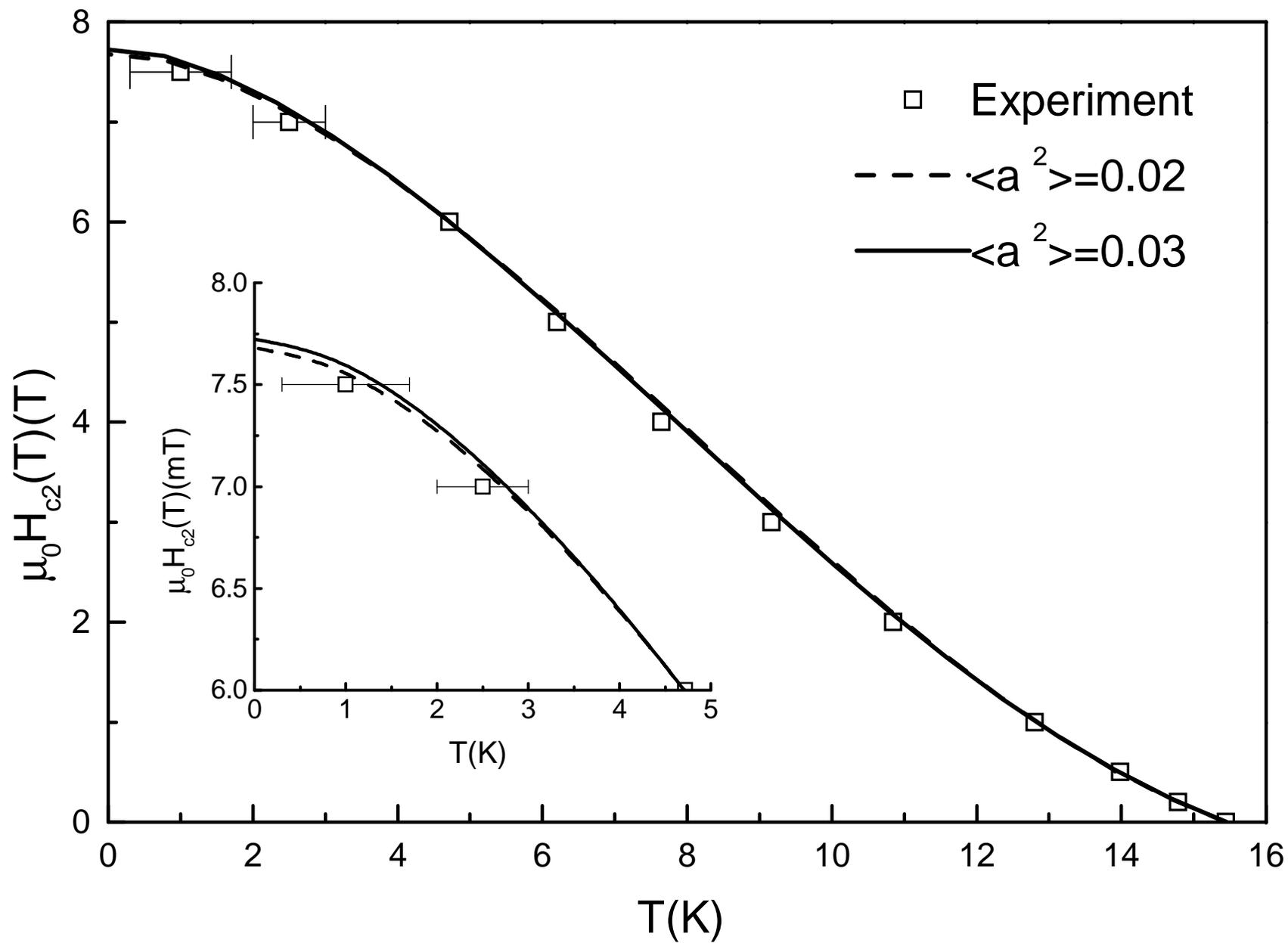
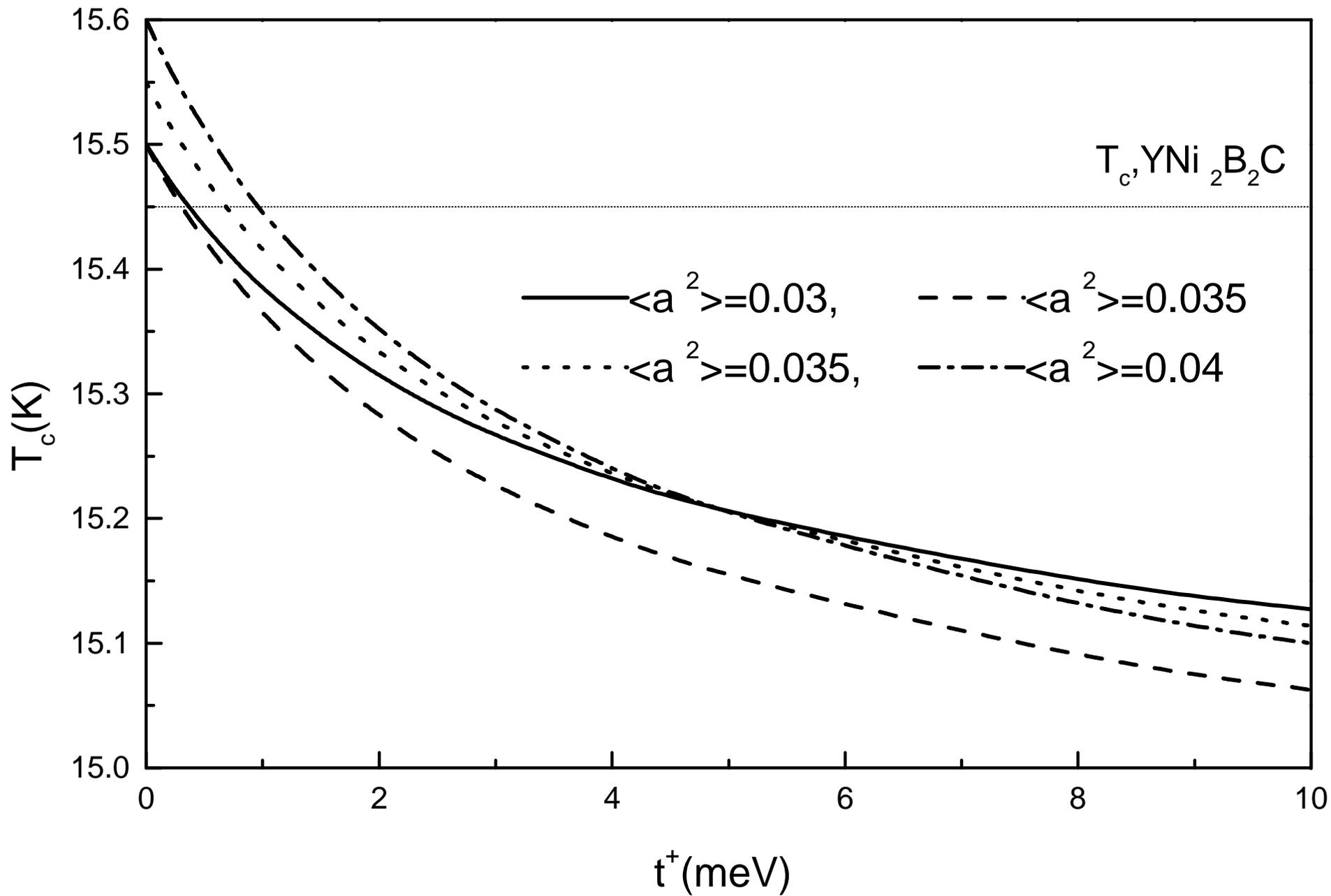
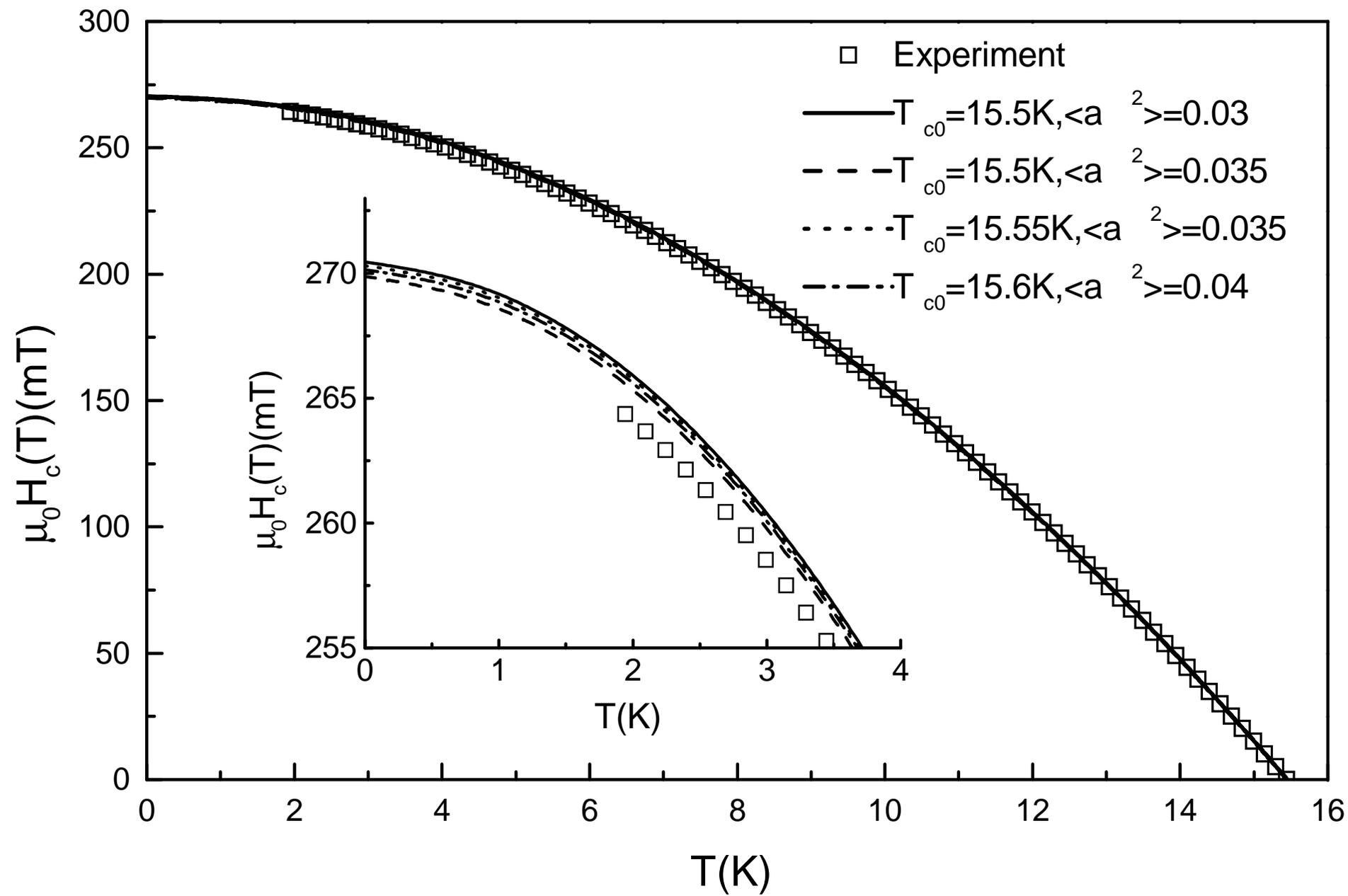
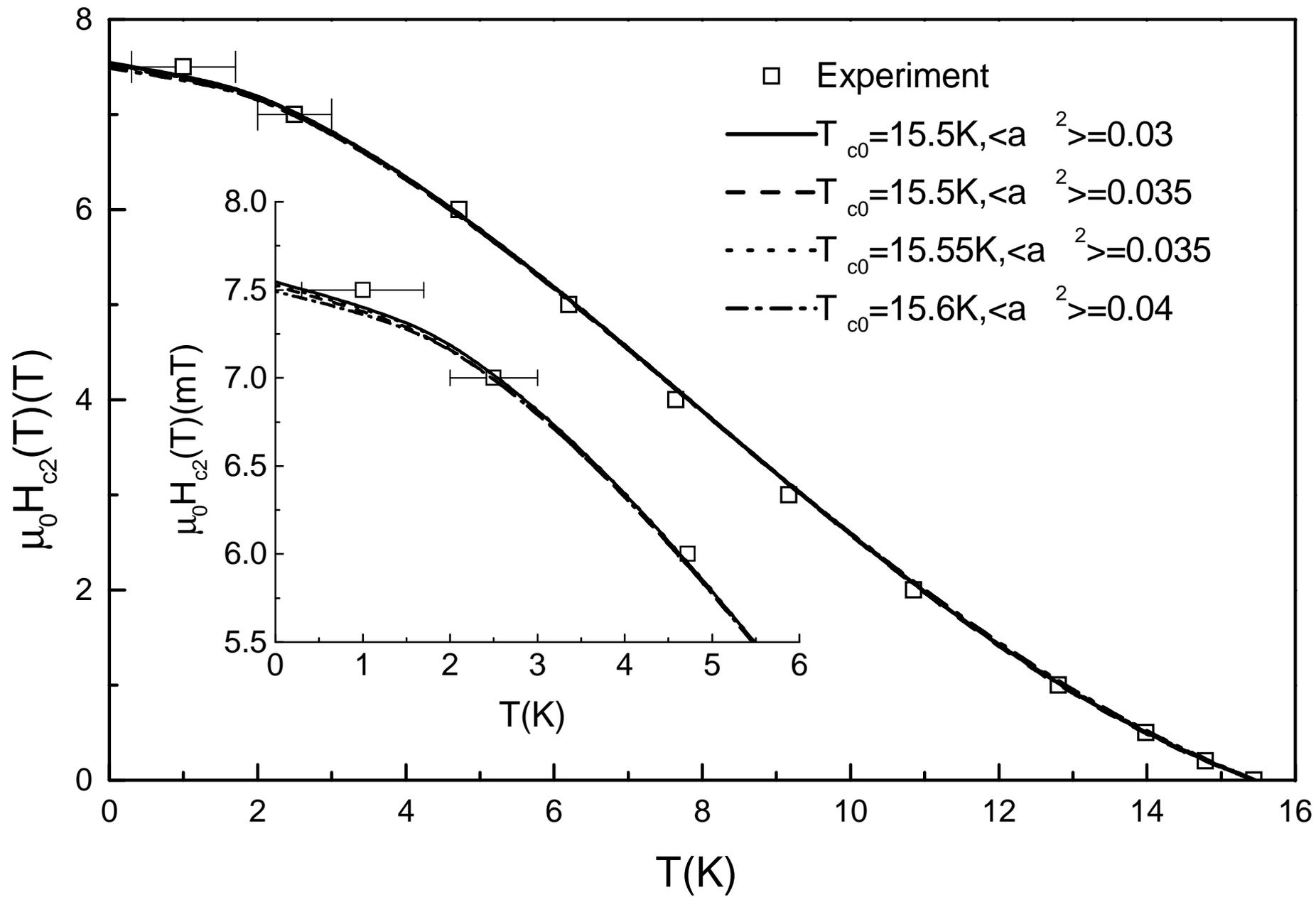
\end{document}